%% file: plc-prov.tex
\documentclass[conference]{IEEEtran}
\PassOptionsToPackage{bookmarks={false}}{hyperref}
\usepackage{graphicx}
\usepackage{url}
\usepackage[pdftex,colorlinks=true,citecolor=black,filecolor=black,%
            linkcolor=black,urlcolor=black]{hyperref}
\usepackage{xspace}

\newcommand{\sys}{{\tt PLC-PROV}\xspace}

\title{Detecting Safety and Security Faults in PLC Systems with Data Provenance}
\author{Abdullah Al Farooq Jessica Marquard, Kripa George, Thomas Moyer\\
UNC Charlotte\\
\{afarooq,jmarqua1,kgeorg16,tom.moyer\}@uncc.edu}
\date{}

\begin{document}

\maketitle

\input{abstract}
\input{intro}
\input{background}

\input{design}
\input{relwork}
\input{conclusion}

\bibliographystyle{IEEEtran}
\bibliography{plcPROVReference}

\end{document}

%% file: abstract.tex
\begin{abstract}
Programmable Logic Controllers are an integral component for managing many different industrial processes (e.g., smart building management, power generation, water and wastewater management, and traffic control systems), and manufacturing and control industries (e.g., oil and natural gas, chemical, pharmaceutical, pulp and paper, food and beverage, automotive, and aerospace). Despite being used widely in many critical infrastructures, PLCs use protocols which make these control systems vulnerable to many common attacks, including man-in-the-middle attacks, denial of service attacks, and memory corruption attacks (e.g., array, stack, and heap overflows, integer overflows, and pointer corruption). In this paper, we propose \sys, a system for tracking the inputs and outputs of the control system to detect violations in the safety and security policies of the system. We consider a smart building as an example of a PLC-based system and show how \sys can be applied to ensure that the inputs and outputs are consistent with the intended safety and security policies.
\end{abstract}

%% file: intro.tex
\section{Introduction} \label{sec:intro}

Industrial control systems rely on automation to ensure safe and efficient operation of many industrial processes including power generation, water and wastewater management, traffic control systems, petroleum and manufacturing industries (e.g., oil and natural gas pipelines, chemical and pharmaceutical production, pulp and paper manufacturing, food and beverage production, and automotive and aerospace assembly). These industries rely heavily on automation to reduce cost and ensure safety of potentially dangerous operations. In the past, these control networks have been isolated from other networks, such as the internet, but we are seeing more and more of these types of networks being connected to the internet~\cite{garcia2017hey,klick2015internet,leszczyna2011protecting}. Often, the goal is to further reduce impact on humans as this new connection enables remote management. However, often the implications of connecting these safety critical control systems is not carefully considered. In many instances, attacks on these systems can lead to disruption of critical services and even loss of human life~\cite{Stouffer2011,interactive1988teen,smith2001hacker,Hancock2003,abrams2007bellingham,falliere2011w32}.

One of the biggest problems with this scenario is that the security of these control networks is limited at best. The communication protocols used in these control networks lack authentication and integrity checking for messages~\cite{Rrushi2012}. These weaknesses make it possible to initiate many commonly-known attacks such as man-in-the-middle attacks, denial of service attacks, memory corruption attacks, replay attacks, and spoofing attacks. While enterprise networks can rely on a wide-range of security mechanisms including IPsec, transport layer security (TLS), and virtual private networks (VPNs) to secure their communications, such mechanisms are difficult to deploy on these control networks, leaving them vulnerable to network-based attackers. Furthermore, many of the commonly applied mitigations fail to cover PLC-based systems~\cite{pincus2004}.

What is needed are mechanisms that can monitor the inputs and outputs of the ICS and ensure that critical safety properties are not violated. This requires an understanding of the desired safety properties, a way to track inputs and outputs, and a mechanism to model the evolution of the system from inputs to outputs. With these mechanisms in place, it becomes possible to ensure that the PLCs do not send commands to actuators that violate the safety and security policies of the system.

In this paper, we propose \sys, a mechanism to track the inputs and outputs of the system and compare them against the specified safety and security properties. \sys relies on tracking \emph{data provenance} for the PLCs and using that provenance to determine if a violation has occurred. Provenance, in short, is the ``history of data transformed by a system'', and has been proposed as a building block for systems that require the ability to reason about the \emph{context} in which an action is taken. Since PLCs are entirely event-driven, context is vitally important, and as such provenance is a natural fit for this sort of analysis.

The rest of this paper is organized as follows. Section~\ref{sec:background} provides relevant background information on PLCs and data provenance. Section~\ref{sec:design} presents the design of \sys and walks through a simple example to highlight how the system works. Finally, Sections~\ref{sec:relwork} and~\ref{sec:conc} detail related work and conclude.

%% file: background.tex
\section{Background} \label{sec:background}
\subsection{Programmable Logic Controllers} \label{secSub:PLC}

\begin{figure*}[ht]
    \centering
    \includegraphics[width=0.7\textwidth]{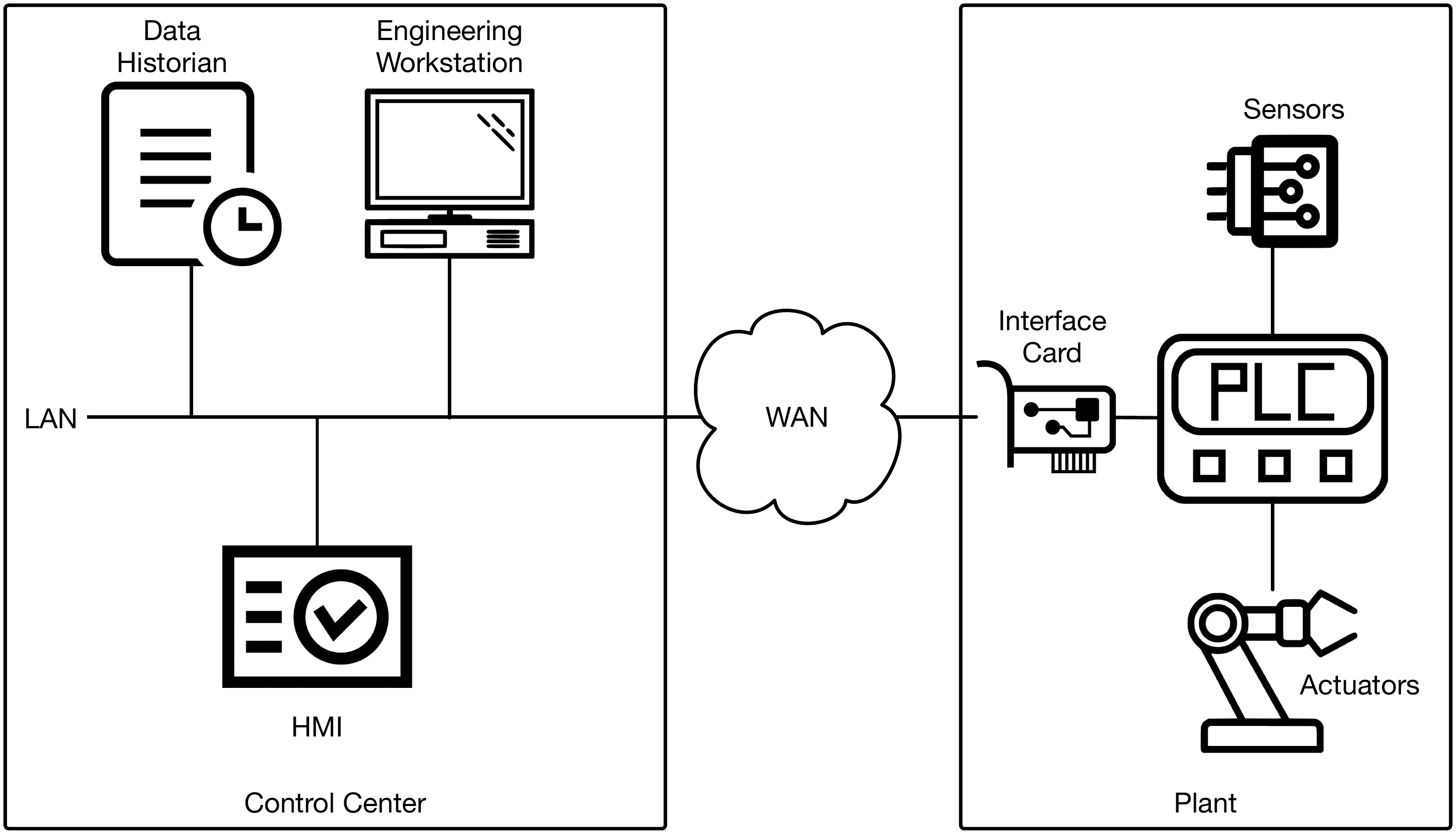}
    \caption{A simple PLC-based control system with the basic components of any industrial control system.}
    \label{PLCsystem}
\end{figure*}

Figure~\ref{PLCsystem} shows a notional Industrial Control System, comprised of several components that together provide the ability to automate industrial processes.
At the heart of this system is the Programmable Logic Controller, or PLC.
The PLC takes input from the sensors and determines the appropriate commands for the actuators to adjust the environment.
The logic for the PLC comes from an engineering workstation that contains the IDE used by the programmer to develop the application logic for the ICS.
Additionally, an ICS has one or more Human Machine Interfaces, or HMIs, that enable operators to view current and historical data from the ICS.
The historical data is stored in the data historian and is often used for post-facto analysis of events.

The PLC applications are written in one of several programming languages including Instruction Lists (IL), Structured Text (ST), Functional Block Diagram (FBD), and Ladder Logic (LL).
Regardless of the programming language used the instructions control the features of the PLC including I/O control, communication, logical decisions, timing, counting, three mode proportional-integral-derivative (PID) control, arithmetic, and data and file processing.
The inputs to the PLC come from a wide array of sensors such as temperature sensors, motion detectors, smoke detectors, water leak detectors, and surveillance cameras.
The outputs of the PLC go to actuators that adjust the current environment.
These actuators include thermostats, humidifiers, speakers, security cameras, video doorbells, door locks, and window blinds.

The topology of an ICS network can be broadly divided into two subnets.
The first is the control network where the sensors and actuators interface with the PLC.
In more complex ICS environments, there may also be a Master Terminal Unit, or MTU, that provides the control programs for the PLCs.
This control network uses non-IP-based protocols such as Modbus~\cite{modbus}.
The second network is the corporate network, where the historian, HMI, and engineering workstation are located.
This is a traditional enterprise network, using standard IP protocols to communicate.
In order to link the corporate and control networks, interface cards are used to provide a bridge between IP-based protocols and the Modbus protocol.
The PLC uses a Modbus/TCP protocol to send data to the historian.
The historian also provides an HTTP interface for devices like the historian to access the stored historical data.

For a distributed system like SCADA (Supervisory control and data acquisition) or DCS (Distributed Control System), a group of PLCs are assigned to different subsystems.
This is mainly done to handle long distance communication among geographically disperse assets (e.g., power grids, natural gas pipelines, water distribution, wastewater collection systems, railway transportation systems).
The far-reaching nature of these systems necessitates numerous control systems responsible for controlling local operations, but working in concert to ensure global functioning of the system.
While these systems are more complex, they rely on many of the same basic components of a smaller-scale ICS.

Even in localized ICS environments (e.g., smart buildings), it is common to rely on several PLCs that work in concert to provide a range of functions.
Consider for example that there might be a PLC that controls the heating, ventilation, and cooling, or HVAC, system, one PLC that controls the elevators, one PLC that controls the door and window locks, and finally one PLC that monitors for hazard conditions (e.g., smoke, carbon monoxide, water and chemical leaks, etc.).
While these systems can be implemented independently, there are often dependencies that need to be accounted for.
Consider a case where the hazard monitoring PLC detects smoke in the building, it must send notifications to the elevator and door/window lock PLCs that this condition is present so that appropriate actions can be taken.
Those actions might be to open the locks on the doors, and move the elevator to the ground floor and then lock out use of the elevator.
These dependencies ensure safety and efficiency in these automated systems.

\subsection{Data Provenance} \label{secSub:DataProvenance}

Data provenance is defined as the ``history of data transformed by a system''~\cite{w3c-prov}.
The provenance of a piece of data describes what the inputs and outputs of each process are, what processes were executed, and who had control of those processes during execution.
Provenance is often represented as a directed acyclic graph (DAG) with nodes representing the data, processes, and controlling entities.
The edges represent causal relationships between these nodes.

Provenance was first introduced in databases and computational sciences for tracing and debugging.
However, more recently it has been proposed as a primitive for building secure and resilient systems~\cite{moyer2016leveraging} that can ``fight through'' attacks.
In order to provide such capabilities, novel collection, storage, and analysis mechanisms have been proposed to enable near-real-time analysis of provenance to support security and resilience decisions~\cite{hahn2018tapp-ids}.
These mechanisms are being used to provide forensic analysis and intrusion detection capabilities~\cite{xie2016unifying}.

%% file: design.tex
\begin{figure}[t]
	\centering
	\includegraphics[scale=0.40, keepaspectratio=true]{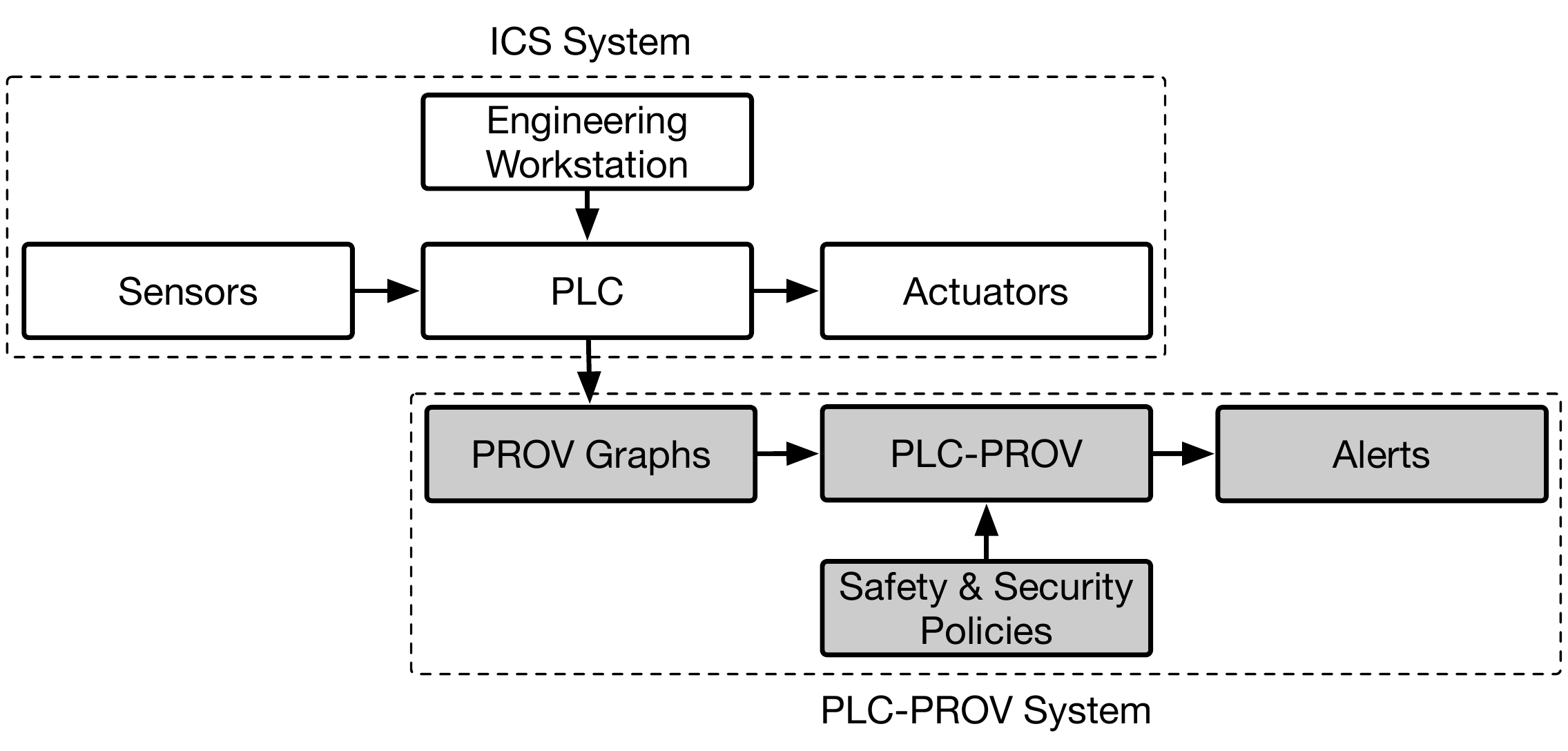}
	\vspace{-9pt}
	\caption{PLC-PROV Architecure}
	\label{Fig_PLCPROV}
	\vspace{-9pt}
\end{figure}

\begin{figure*}[ht]
	\centering
	\includegraphics[width=.8\textwidth, keepaspectratio=true]{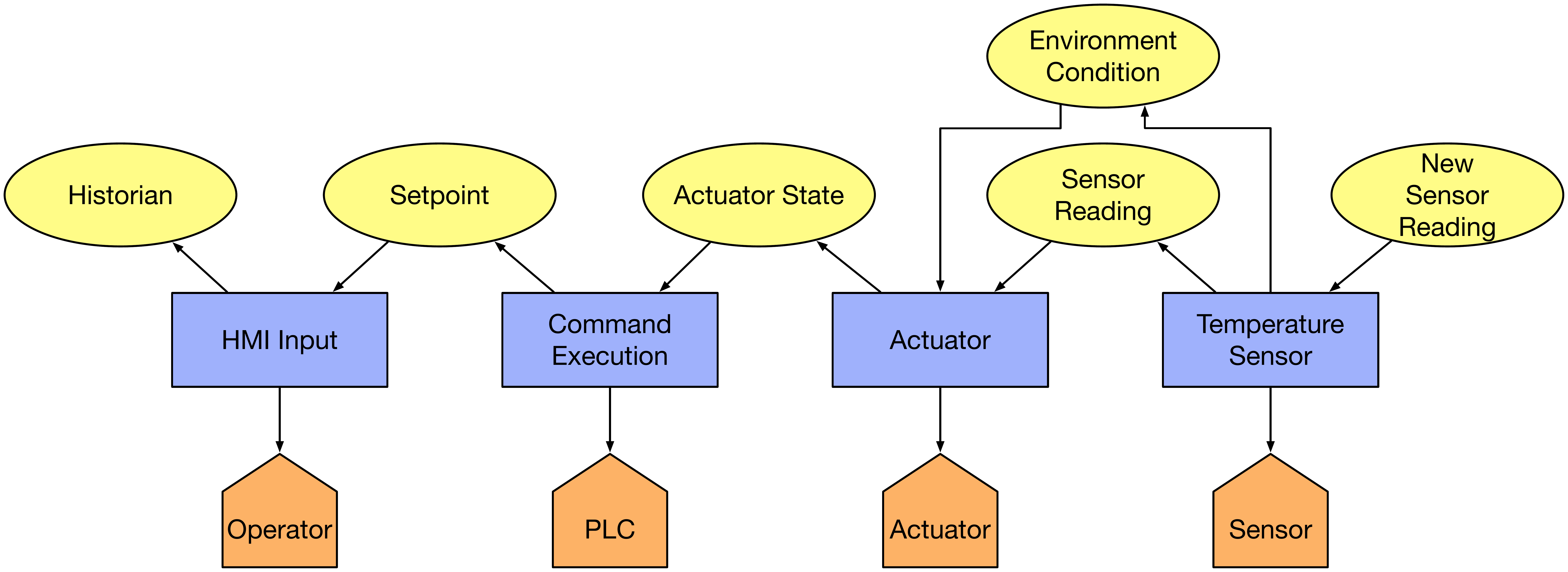}
	\vspace{-9pt}
	\caption{Provenance Model for PLC-based System}
	\label{Fig_GraphModel}
	\vspace{-9pt}
\end{figure*}

\section{Design} \label{sec:design}

Due to the distributed nature of PLC systems an attacker can trigger an event that leads to conflicting actions for the same object or feature of the plant/environment.
Let us consider a smart building as an example where PLC is used \cite{kastner2005communication,granzer2010security,barz2016plcs,sysala2016monitoring,skeledzija2014smart}.
An attacker can compromise a carbon monoxide detector and trigger a false alarm.
This sends a command to the smart windows to open allowing fresh air into the building.
Occupants will evacuate the building and a thief can use the open windows to enter the building.
Another example can be creating multiple events that trigger a thermostat to increase and decrease the temperature of a room at the same time.
Sending two different commands in the thermostat at the same time continuously can damage it, by artificially shortening the device's lifespan.
In this way, the attacker not only damages an asset but also may drive the occupants of the room to leave due to fluctuations in the comfort level of the room.
In addition to the attacks described above, an attacker can create a series of attacks (cascading attack)~\cite{IoTDDOS}.
Moreover, misconfiguration in the smart building operation is possible as there are numerous rules or policies for taking actions by the controllers after events have occurred.

A formal methods approach for detecting conflicts in IoT system is presented in \cite{IoTC2}.
A PLC-based system (e.g., smart home, power supply, water supply, wastewater management, and traffic control system) works on the same sensor-actuator-controller functionalities.
Therefore, we adopted the safety policies defined in \cite{IoTC2} as the basic policies to analyze using the collected provenance graphs.
\sys will check whether there is any violation of the defined safety and security properties.
If found, it is reported as an anomaly for the system, which also provides the administrator with the detailed traces that are needed to understand the impact of the anomaly.
The basic architecture is given in Figure \ref{Fig_PLCPROV}.

The PLC contains the core rules/logic for controlling the plant/environment.
The controller issues command to the actuators based on these sensor measurements.
A  PLC has to be operated with software that provides interaction with the sensors and actuators.
These addresses are mapped as variables in the source program.
In this project, we use CODESYS\footnote{\url{https://www.codesys.com/}} which is a development system for PLC applications.
To start, our framework traces the variables designated for sensors and actuators.
These values of these variables are collected with timestamps into traces of system execution.

These traces are the input for a provenance management tool \emph{Curator}~\cite{curator}.
Curator is a lightweight library which minimizes integration complexity for the application developers.
It is also capable of integrating provenance from multiple levels of abstraction, a feature that enables reasonsing about provenance both at the sensor reading level (micro) and at the environment/plant level (macro).
As Curator is targeted for microservice based system, it fits well for our case where we collect traces from disparate components of the system, similar to microservices.
From there, the provenance graph is generated, showing the evolution of the system from sensor readings through the PLC and controller, to the actuators.
Figure~\ref{Fig_GraphModel} is a notional example of the type of graph that is generated.

The provenance graphs collected by \sys enable an administrator to answer the following questions:
\begin{itemize}
	\item Has an actuator been actuated more than once at the same time?
	\item If an actuator receives multiple commands at the same time, are these same or different?
	\item What are the reasons behind conflicting action?
	\item Which are the sensors influencing conflicting commands?
	\item Has any sensor measurement gone beyond normal range? If yes, how many times did that happen and how long did it last?
	\item Are there more than two actions affecting the same environment feature?	
\end{itemize}

\subsection{Example}

\begin{figure*}[ht]
	\centering
	\includegraphics[width=.9\textwidth, keepaspectratio=true]{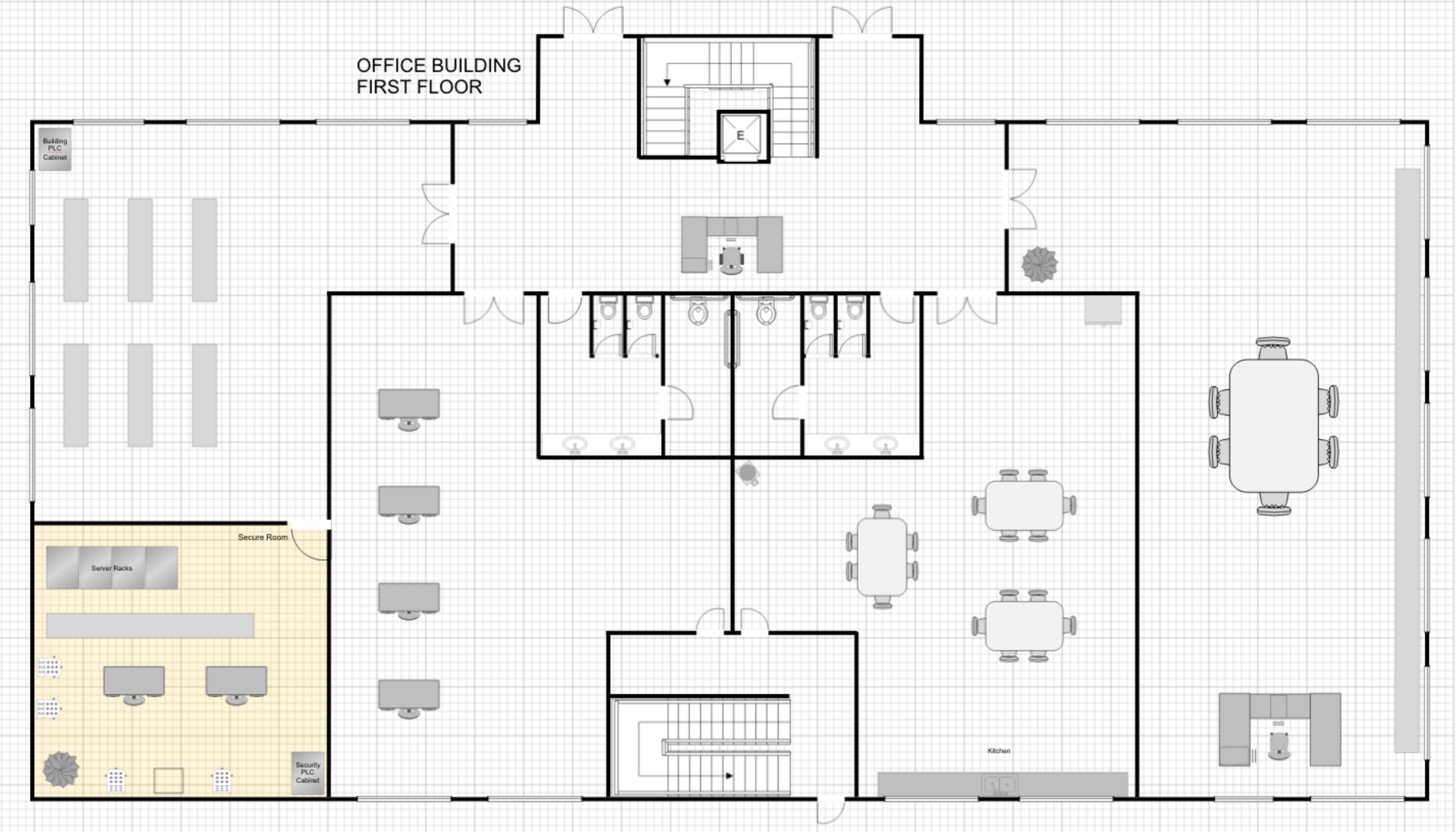}
	\caption{Notional office floorplan with a secure area that is access controlled and environmentally monitored.}
	\label{fig:notional-floorplan}
\end{figure*}

In order to illustrate how \sys will operate, we provide a short example of an attack scenario.
Consider the floorplan shown in Figure~\ref{fig:notional-floorplan} which includes a secure, access-controlled area in the lower-left corner.
In this scenario, we consider the policy that security doors should only be unlocked when presented with a valid access card and in an emergency when human life is at stake.
In our scenario, we have PLCs controlling various aspects of our smart building.
The first we call our \emph{safety PLC} which contains a smoke detector as an input.
The output of this PLC is connected to an alarm system and to our \emph{security PLC}.
The security PLC has a card reader as input and a lock control on its output.
Finally, we have the \emph{environmental PLC} which monitors environmental conditions (such as temperature and humidity).
As a safety mechanism, when hazards are detected that require evacuation, security doors should be unlocked to ensure no one is locked in the building, also allowing first responders to clear all areas.

Given this setup, we now consider an attacker that wishes to gain physical access to the secured areas within the building.
In order to gain access, the attacker first poses as a maintenance technician to gain access to the common areas of the building.
From here, he connects to the PLC network and forges a sensor reading from the smoke alarm indicating a hazard condition in the secure area.
This hazard condition triggers the alarm to evacuate the secure area, which also causes the doors to unlock.
In this scenario, the attacker can use the unlock and evacuate behavior to gain access to the secure area.

\sys aids the adminstrators in detecting this attack in the following way.
As it monitors the flow of inputs and outputs of the system, it would take into account context.
There are several indicators that something is likely wrong.
One, smoke detector alarms will typically correspond to a rise in temperature as well, which won't be present in the attack scenario.
If the attacker is smart enough, he can simulate this behavior.
However, another indication of malicious behavior is that the smoke detector signal and temperature readings would have to be present in the secure space and that means that the smoke detector reading at the PLC would be on a specific input line.
However, since the attacker doesn't have access to the space, they have to fake the measurement from another point within the control network.
Here we assume that the PLCs controlling the secured space are located within the secured space itself, as is typical practice in building secure areas, while the environmental PLCs are located in the common areas of the building to allow technicians to gain access. This is depicted in Figure~\ref{fig:notional-floorplan} with the upper-left block representing the enivonmental PLC cabinet and the lower-right block in the secure area representing the security PLC cabinet.
It is also assumed that in order for the alarm from the environmental sensors to reach the secure area that there is a connection between the two cabinets.
By monitoring the flow of inputs and outputs, \sys would be able to detect this deviation in known good behavior and flag it as anomalous, further alerting building personnel to the potential issue.
In this way, we can ensure that safety and security policies are not violated.

%% file: relwork.tex
\section{Related Work} \label{sec:relwork}
The closest work to \sys is PROV-CPS~\cite{nwafor2018trace} where provenance was collected from resource-constrained embedded devices of the cyber-physical system.
However, this research collects provenance from sensors only to identify anomalous measurements.
On the contrary, apart from collecting provenance from the sensor measurement, our work covers the actions of the actuators and the dependencies among the PLCs in finding malicious activities.
Our approach is complete in expressing the causality and dependencies among the data objects through the provenance graph.
Another notable work in this area is ProvThings~\cite{wang2018fear} where a provenance collection framework is proposed for IoT apps and devices.
ProvThings presents an automated instrumentation mechanism for IoT apps and device APIs.
The collected provenance is then used to generate explanations for ``why'' a particular action occurred.
Our work captures provenance data for all sensors and actuators in order to detect safety and security policy violations.

There have been several other attempts to deploy security policies with static verification, dynamic verification, and the hybrid of these two approaches.
Static verification (model checking) is proposed in TSV~\cite{mclaughlin2014trusted} where a middleware ensures the safety of a PLC-based system sitting between PLC and the devices.
TSV verifies the safety behavior of the code executed on PLC before commands reach the actuators.
The safety properties are written in temporal logic which is verified using model checking.
While this work verifies the system's behavior, there are some other works that started the verification from PLCs' source program~\cite{sadolewski2011automated,sadolewski2011conversion}.
Later, others proposed mechanism to automatically generate formal models from PLC programs~\cite{biallas2012arcade,darvas2013transforming,adiego2014bringing,markovic2015automated,enoiu2016automated}.

While static analysis performs the verification before the PLC program is released for operation (i.e. compile-time), dynamic analysis ensures that policies are not violated at run-time.
$C^2$~\cite{mclaughlin2013cps} introduced an enforcement mechanism for safety policies in PLC-based system.
When a PLC issues a command to an actuator, the current states of the system are checked and then decisions are made whether or not the command should be issued through their enforcement mechanism, $C^2$.
In this work, concerns about the size of the trusted computing base (TCB) and state explosion in the model checking were expressed.
\cite{mclaughlin2015blocking} addresses reduces the size of the TCB considerably, and merges the static and dynamic analysis of TSV and $C^2$.
The works by McLaughlin, et. al. focus specifically on safety properties.
This was subsequently extended in~\cite{zonouz2014detecting} with an effort to find malicious PLC programs.
Another approach to dynamic analysis is proposed in~\cite{nicholson2014position} using Interval Temporal Logic (ITL) and the Tempura framework, which aims to provide early alerts in PLC-based systems.

Later, this work was extended in~\cite{janicke2015runtime} where an ITL/Tempura definition of a Siemens S7-1200 PLC ladder logic was presented.
Their developed monitoring methodology captures a snapshot of the current state (with values for markers, input, output, counters, and timers) of the PLC.
Tempura was implemented to execute on an Arduino Uno connected to the PLC, ensuring that the PLC does not need a powerful computing node to perform the computations.

While static analysis has proven promising, the number of possible inputs and outputs for a PLC system can lead to a state explosion.
Furthermore, dynamic analysis suffers from a coverage problem, where only executed code paths are verified.
Symbolic execution can minimize the state space, but cannot guarantee complete verification of outputs (actuation command) from input sets (sensor measurement).
For these reasons, what is needed is a mechanism that can provide high-level safety and security policy descriptions that can be enforced at run-time where the appropriate context can be considered.

%% file: conclusion.tex
\section{Conclusion} \label{sec:conc}
This project focuses on the integration of data provenance in PLC controlled system in order to detect safety policy violation there. We have modeled data provenance that considers operator (through HMI) input, command execution, actuators' state, and sensors' measurement. Therefore, we claim that our model is complete in expressing the causality and dependencies among the data objects in a PLC-controlled system. We evaluated our model with smart building environment. It turns out that data provenance has great potential applicability in PLC controlled system where the change of sensor measurement and actuator actions take place very frequently. Despite being used in critical infrastructures, PLCs have little or almost no security. The integration of plc-prov is capable of enforcing adequate safety and security policies. 

%% file: plc-prov.bbl
\begin{thebibliography}{}
\providecommand{\url}[1]{#1}
\csname url@samestyle\endcsname
\providecommand{\newblock}{\relax}
\providecommand{\bibinfo}[2]{#2}
\providecommand{\BIBentrySTDinterwordspacing}{\spaceskip=0pt\relax}
\providecommand{\BIBentryALTinterwordstretchfactor}{4}
\providecommand{\BIBentryALTinterwordspacing}{\spaceskip=\fontdimen2\font plus
\BIBentryALTinterwordstretchfactor\fontdimen3\font minus
  \fontdimen4\font\relax}
\providecommand{\BIBforeignlanguage}[2]{{%
\expandafter\ifx\csname l@#1\endcsname\relax
\typeout{** WARNING: IEEEtran.bst: No hyphenation pattern has been}%
\typeout{** loaded for the language `#1'. Using the pattern for}%
\typeout{** the default language instead.}%
\else
\language=\csname l@#1\endcsname
\fi
#2}}
\providecommand{\BIBdecl}{\relax}
\BIBdecl

\end{thebibliography}


\begin{thebibliography}{10}
\providecommand{\url}[1]{#1}
\csname url@samestyle\endcsname
\providecommand{\newblock}{\relax}
\providecommand{\bibinfo}[2]{#2}
\providecommand{\BIBentrySTDinterwordspacing}{\spaceskip=0pt\relax}
\providecommand{\BIBentryALTinterwordstretchfactor}{4}
\providecommand{\BIBentryALTinterwordspacing}{\spaceskip=\fontdimen2\font plus
\BIBentryALTinterwordstretchfactor\fontdimen3\font minus
  \fontdimen4\font\relax}
\providecommand{\BIBforeignlanguage}[2]{{%
\expandafter\ifx\csname l@#1\endcsname\relax
\typeout{** WARNING: IEEEtran.bst: No hyphenation pattern has been}%
\typeout{** loaded for the language `#1'. Using the pattern for}%
\typeout{** the default language instead.}%
\else
\language=\csname l@#1\endcsname
\fi
#2}}
\providecommand{\BIBdecl}{\relax}
\BIBdecl

\bibitem{garcia2017hey}
L.~Garcia, F.~Brasser, M.~H. Cintuglu, A.-R. Sadeghi, O.~Mohammed, and S.~A.
  Zonouz, ``Hey, my malware knows physics attacking plcs with physical model
  aware rootkit,'' in \emph{Proceedings of the Network \& Distributed System
  Security Symposium, San Diego, CA, USA}, 2017, pp. 26--28.

\bibitem{klick2015internet}
J.~Klick, S.~Lau, D.~Marzin, J.-O. Malchow, and V.~Roth, ``Internet-facing plcs
  as a network backdoor,'' in \emph{Communications and Network Security (CNS),
  2015 IEEE Conference on}.\hskip 1em plus 0.5em minus 0.4em\relax IEEE, 2015,
  pp. 524--532.

\bibitem{leszczyna2011protecting}
R.~Leszczyna, E.~Egozcue, L.~Tarrafeta, V.~F. Villar, R.~Estremera, and
  J.~Alonso, ``{Protecting industrial control systems-recommendations for
  europe and member states},'' European Union Agency for Network and
  Information Security,
  \url{https://www.enisa.europa.eu/publications/protecting-industrial-control-systems.-recommendations-for-europe-and-member-states},
  Tech. Rep., 2011.

\bibitem{Stouffer2011}
K.~Stouffer, J.~Falco, and K.~Scarfone, ``Guide to industrial control systems
  (ics) security,'' \emph{NIST special publication}, vol. 800, no.~82, pp.
  16--16, 2011.

\bibitem{interactive1988teen}
\BIBentryALTinterwordspacing
C.~Interactive, ````teen hacker faces federal charges,'' \emph{viewed on},
  vol.~15, 1988. [Online]. Available:
  \url{http://www.cnn.com/TECH/computing/9803/18/juvenile.hacker/index.html}
\BIBentrySTDinterwordspacing

\bibitem{smith2001hacker}
\BIBentryALTinterwordspacing
T.~Smith, ``Hacker jailed for revenge sewage attacks,'' \emph{The Register},
  vol.~31, 2001. [Online]. Available:
  \url{https://www.theregister.co.uk/2001/10/31/hacker_jailed_for_revenge_sewage/}
\BIBentrySTDinterwordspacing

\bibitem{Hancock2003}
D.~Hancock, ``Virus disrupts train signals,'' Aug. 2003.

\bibitem{abrams2007bellingham}
M.~Abrams and J.~Weiss, ``Bellingham control system cyber security case
  study,'' 2007.

\bibitem{falliere2011w32}
N.~Falliere, L.~O. Murchu, and E.~Chien, ``W32. stuxnet dossier,'' \emph{White
  paper, Symantec Corp., Security Response}, vol.~5, no.~6, p.~29, 2011.

\bibitem{Rrushi2012}
\BIBentryALTinterwordspacing
J.~L. Rrushi, \emph{SCADA Protocol Vulnerabilities}.\hskip 1em plus 0.5em minus
  0.4em\relax Berlin, Heidelberg: Springer Berlin Heidelberg, 2012, pp.
  150--176. [Online]. Available:
  \url{https://doi.org/10.1007/978-3-642-28920-0_8}
\BIBentrySTDinterwordspacing

\bibitem{pincus2004}
J.~Pincus and B.~Baker, ``Mitigations for low-level coding vulnerabilities:
  Incomparability and limitations.''

\bibitem{modbus}
{Modbus Organization}, ``{Modbus Application Protocol Specification V1.1b3},''
  \url{http://www.modbus.org/docs/Modbus_Application_Protocol_V1_1b3.pdf}.

\bibitem{w3c-prov}
{World Wide Web Consortium and others}, ``{PROV-Overview: an overview of the
  PROV family of documents},'' \url{https://www.w3.org/TR/prov-overview/},
  2013.

\bibitem{moyer2016leveraging}
T.~Moyer, K.~Chadha, R.~Cunningham, N.~Schear, W.~Smith, A.~Bates, K.~Butler,
  F.~Capobianco, T.~Jaeger, and P.~Cable, ``Leveraging data provenance to
  enhance cyber resilience,'' in \emph{Cybersecurity Development (SecDev),
  IEEE}.\hskip 1em plus 0.5em minus 0.4em\relax IEEE, 2016, pp. 107--114.

\bibitem{hahn2018tapp-ids}
\BIBentryALTinterwordspacing
X.~Han, T.~Pasquier, and M.~Seltzer, ``Provenance-based intrusion detection:
  Opportunities and challenges,'' in \emph{10th {USENIX} Workshop on the Theory
  and Practice of Provenance (TaPP 2018)}.\hskip 1em plus 0.5em minus
  0.4em\relax London: {USENIX} Association, 2018. [Online]. Available:
  \url{https://www.usenix.org/conference/tapp2018/presentation/han}
\BIBentrySTDinterwordspacing

\bibitem{xie2016unifying}
Y.~Xie, D.~Feng, Z.~Tan, and J.~Zhou, ``Unifying intrusion detection and
  forensic analysis via provenance awareness,'' \emph{Future Generation
  Computer Systems}, vol.~61, pp. 26--36, 2016.

\bibitem{kastner2005communication}
W.~Kastner, G.~Neugschwandtner, S.~Soucek, and H.~M. Newman, ``Communication
  systems for building automation and control,'' \emph{Proceedings of the
  IEEE}, vol.~93, no.~6, pp. 1178--1203, 2005.

\bibitem{granzer2010security}
W.~Granzer, F.~Praus, and W.~Kastner, ``Security in building automation
  systems,'' \emph{IEEE Transactions on Industrial Electronics}, vol.~57,
  no.~11, pp. 3622--3630, 2010.

\bibitem{barz2016plcs}
C.~Barz, S.~Deaconu, T.~Latinovic, A.~Berdie, A.~Pop-Vadean, and M.~Horgos,
  ``Plcs used in smart home control,'' in \emph{IOP Conference Series:
  Materials Science and Engineering}, vol. 106, no.~1.\hskip 1em plus 0.5em
  minus 0.4em\relax IOP Publishing, 2016, p. 012036.

\bibitem{sysala2016monitoring}
T.~Sysala, M.~Posp{\'\i}chal, and P.~Neumann, ``Monitoring and control system
  for a smart family house controlled via programmable controller,'' in
  \emph{Carpathian Control Conference (ICCC), 2016 17th International}.\hskip
  1em plus 0.5em minus 0.4em\relax IEEE, 2016, pp. 706--710.

\bibitem{skeledzija2014smart}
N.~Skeledzija, J.~Cesic, E.~Koco, V.~Bachler, H.~N. Vucemilo, and H.~Dzapo,
  ``Smart home automation system for energy efficient housing,'' in
  \emph{Information and Communication Technology, Electronics and
  Microelectronics (MIPRO), 2014 37th International Convention on}.\hskip 1em
  plus 0.5em minus 0.4em\relax IEEE, 2014, pp. 166--171.

\bibitem{IoTDDOS}
\BIBentryALTinterwordspacing
S.~COBB, ``10 things to know about the october 21 iot ddos attacks,'' 2016.
  [Online]. Available:
  \url{http://www.welivesecurity.com/2016/10/24/10-things-know-october-21-iot-ddos-attacks/}
\BIBentrySTDinterwordspacing

\bibitem{IoTC2}
A.~A. Farooq, E.~Al-Shaer, T.~Moyer, and K.~Kant, ``Iotc2: A formal method
  approach for detecting conflicts in large scale iot systems,'' in
  \emph{Integrated Network and Service Management (IM), 2017 IFIP/IEEE
  Symposium on}.\hskip 1em plus 0.5em minus 0.4em\relax IEEE, 2019.

\bibitem{curator}
\BIBentryALTinterwordspacing
W.~Smith, T.~Moyer, and C.~Munson, ``Curator: Provenance management for modern
  distributed systems,'' \emph{CoRR}, vol. abs/1806.02227, 2018. [Online].
  Available: \url{http://arxiv.org/abs/1806.02227}
\BIBentrySTDinterwordspacing

\bibitem{nwafor2018trace}
E.~Nwafor, ``Trace-based data provenance for cyber-physical systems,'' Ph.D.
  dissertation, Howard University, 2018.

\bibitem{wang2018fear}
Q.~Wang, W.~U. Hassan, A.~Bates, and C.~Gunter, ``Fear and logging in the
  internet of things,'' in \emph{ISOC NDSS}, 2018.

\bibitem{mclaughlin2014trusted}
S.~E. McLaughlin, S.~A. Zonouz, D.~J. Pohly, and P.~D. McDaniel, ``A trusted
  safety verifier for process controller code.'' in \emph{NDSS}, vol.~14, 2014.

\bibitem{sadolewski2011automated}
J.~Sadolewski, ``Automated conversion of st control programs to why for
  verification purposes,'' in \emph{Computer Science and Information Systems
  (FedCSIS), 2011 Federated Conference on}.\hskip 1em plus 0.5em minus
  0.4em\relax IEEE, 2011, pp. 849--854.

\bibitem{sadolewski2011conversion}
------, ``Conversion of st control programs to ansi c for verification
  purposes.'' \emph{e-Informatica}, vol.~5, no.~1, pp. 65--76, 2011.

\bibitem{biallas2012arcade}
S.~Biallas, J.~Brauer, and S.~Kowalewski, ``Arcade. plc: A verification
  platform for programmable logic controllers,'' in \emph{Proceedings of the
  27th IEEE/ACM International Conference on Automated Software
  Engineering}.\hskip 1em plus 0.5em minus 0.4em\relax ACM, 2012, pp. 338--341.

\bibitem{darvas2013transforming}
\BIBentryALTinterwordspacing
D.~Darvas, B.~Fernandez~Adiego, and E.~Blanco, ``{Transforming PLC Programs
  into Formal Models for Verification Purposes},'' {CERN}, Tech. Rep., Nov
  2013. [Online]. Available: \url{http://cds.cern.ch/record/1629275}
\BIBentrySTDinterwordspacing

\bibitem{adiego2014bringing}
B.~F. Adiego, D.~Darvas, J.-C. Tournier, E.~B. Vinuela, and V.~M.~G.
  Su{\'a}rez, ``Bringing automated model checking to plc program
  development---a cern case study---,'' \emph{IFAC Proceedings Volumes},
  vol.~47, no.~2, pp. 394--399, 2014.

\bibitem{markovic2015automated}
F.~Markovic, ``Automated test generation for structured text language using
  uppaal model checker,'' 2015.

\bibitem{enoiu2016automated}
E.~P. Enoiu, A.~{\v{C}}au{\v{s}}evi{\'c}, T.~J. Ostrand, E.~J. Weyuker,
  D.~Sundmark, and P.~Pettersson, ``Automated test generation using model
  checking: an industrial evaluation,'' \emph{International Journal on Software
  Tools for Technology Transfer}, vol.~18, no.~3, pp. 335--353, 2016.

\bibitem{mclaughlin2013cps}
S.~McLaughlin, ``Cps: Stateful policy enforcement for control system device
  usage,'' in \emph{Proceedings of the 29th Annual Computer Security
  Applications Conference}.\hskip 1em plus 0.5em minus 0.4em\relax ACM, 2013,
  pp. 109--118.

\bibitem{mclaughlin2015blocking}
------, ``Blocking unsafe behaviors in control systems through static and
  dynamic policy enforcement,'' in \emph{Design Automation Conference (DAC),
  2015 52nd ACM/EDAC/IEEE}.\hskip 1em plus 0.5em minus 0.4em\relax IEEE, 2015,
  pp. 1--6.

\bibitem{zonouz2014detecting}
S.~Zonouz, J.~Rrushi, and S.~McLaughlin, ``Detecting industrial control malware
  using automated plc code analytics,'' \emph{IEEE Security \& Privacy},
  vol.~12, no.~6, pp. 40--47, 2014.

\bibitem{nicholson2014position}
A.~Nicholson, H.~Janicke, and A.~Cau, ``Position paper: Safety and security
  monitoring in ics/scada systems.'' in \emph{ICS-CSR}, 2014.

\bibitem{janicke2015runtime}
H.~Janicke, A.~Nicholson, S.~Webber, and A.~Cau, ``Runtime-monitoring for
  industrial control systems,'' \emph{Electronics}, vol.~4, no.~4, pp.
  995--1017, 2015.

\end{thebibliography}
